\documentclass[fleqn,usenatbib]{mnras}
\usepackage{amsmath}
\usepackage[utf8]{inputenc}
\usepackage{graphicx}
\usepackage{txfonts}
\usepackage[para,online,flushleft]{threeparttable}
\usepackage{xcolor}
\usepackage{xfrac}
\usepackage{subcaption}
\definecolor{redak}{rgb}{0.9,0.15,0.05}

\usepackage{svg}
\newcommand{\orcid}[1]{\href{https://orcid.org/#1}{\includesvg[width=10pt]{orcid}}}

\usepackage{multirow} 

\usepackage{xspace}
\newcommand{\mesa}{\textsc{mesa}\xspace} 

\usepackage{tikz}
\usetikzlibrary{shapes.geometric, arrows.meta, positioning, calc}

\defcitealias{sander_nature_2020}{S20}
\defcitealias{sander_temperature_2023}{S23}

\title[WR evolution with temperature-dependent winds]{The formation and evolution of Wolf--Rayet stars with temperature-dependent winds}
\author[Flint et al.]{Guy~O.~Flint$^{\orcid{0009-0006-9408-6029}}$, $^{1, 2}$\thanks{\href{guy.flint@physics.ox.ac.uk}{guy.flint@physics.ox.ac.uk}} Avishai~Gilkis$^{\orcid{0000-0001-8949-5131}}$, $^{1,3}$\thanks{\href{a.gilkis@surrey.ac.uk}{a.gilkis@surrey.ac.uk}} Tomer~Shenar$^{\orcid{0000-0003-0642-8107}}$ $^{4}$ and Christopher~A.~Tout$^{\orcid{0000-0002-1556-9449}}$ $^{1}$ \\
$^{1}$ Institute of Astronomy, University of Cambridge, Madingley Road, Cambridge CB3 0HA, United Kingdom\\
$^{2}$ Sub-department of Astrophysics, Department of Physics, University of Oxford, Denys Wilkinson Building, Keble Road, Oxford OX1 3RH, United Kingdom\\
$^{3}$ School of Mathematics and Physics, University of Surrey, Stag Hill, Guildford GU2 7XH, United Kingdom\\
$^{4}$ The School of Physics and Astronomy, Tel Aviv University, Tel Aviv 6997801, Israel}

\pubyear{2026}

\begin{document}
\label{firstpage}
\pagerange{\pageref{firstpage}--\pageref{lastpage}}

\maketitle

\begin{abstract}
Massive stars strongly influence their environments through their radiation, stellar winds and core-collapse supernovae, while their own evolution is strongly shaped by wind mass loss. Wolf--Rayet (WR) stars are characterised by strong, optically thick winds, and the implementation of these winds remains an important uncertainty in stellar evolution models. Recent hydrodynamically consistent atmosphere calculations predict WR mass-loss rates that depend strongly on the temperature at the wind-launching region. We investigate the effects of incorporating these latest theoretical WR wind prescriptions into single and binary stellar evolution models at Milky Way, Large Magellanic Cloud and Small Magellanic Cloud (SMC) metallicities. We find that the temperature dependence of the WR wind can significantly alter the evolution of stars near the threshold for WR formation. In particular, an implementation that transitions from a hydrogen-rich WNL wind prescription to a classical WR wind driven by the hot iron opacity bump causes some SMC-metallicity models to spend an extended period at temperatures of approximately $70$ to $90\,\mathrm{kK}$.
This produces a concentration of models in the temperature range occupied by most observed SMC WR stars, a feature that has previously been difficult to explain with stellar evolution models. 
Our results demonstrate that incorporating physically motivated, temperature-dependent WR winds can improve the agreement between stellar evolution models and observed WR populations.
\end{abstract}


\section{Introduction}
\label{sec:intro}



Massive stars, those with an initial mass $M\gtrsim 8\,\mathrm{M}_\odot$, play a crucial role in a range of astrophysical processes. Despite their rarity, they dominate the ionising and mechanical feedback of the interstellar medium (ISM) through their high luminosities and strong stellar winds (\citealt*{smith_realistic_2002}; \citealt{robertson_early_2010, krause_feedback_2013}). They further enrich the ISM chemically and energetically through core-collapse supernovae (CCSNe) at the end of their lives. Neutron stars (NSs) and black holes (BHs) are the remnants of these massive stars \citep{heger_how_2003, ertl_explosion_2020} and, in binary systems, they can become sources of gravitational waves originating from their coalescence \citep[e.g.][]{abbott_gw151226_2016}.

Unlike for their low-mass counterparts, wind mass loss from massive stars governs their evolution and affects their spectral appearances throughout their evolution. Notably, classical Wolf--Rayet (cWR) stars form as a late stage in the life of massive stars when they have completely lost their outer hydrogen layers and are fusing helium or heavier elements in their cores. Their strong, optically thick winds give rise to their characteristic prominent, broad emission lines, which are observable even in integrated spectra of entire galaxies \citep*{conti_wolfrayet_1991, schaerer_populations_1998}. WR classifications are based on the relative strengths of their spectral lines: WN, WC and WO stars have prominent nitrogen, carbon and oxygen emission lines, respectively \citep{smith_revised_1968, sander_wolfrayet_2014}. This may be viewed as an evolutionary sequence. As the star becomes increasingly stripped and evolved its surface becomes N-rich through CNO-cycle processing and then CO-rich from helium burning \citep{crowther_physical_2007, shenar_wolf_2019}. However, such a sequence is not guaranteed before the end of the life of a star. A precursor to a WN star may fall under the hydrogen-rich WNh classification. Such stars have some remaining surface hydrogen and may still be main-sequence (MS) core-hydrogen burning stars (\citealt*{deKoter1997,hamann_galactic_2006}; \citealt{grafener_eddington_2011}).

WR winds are understood to be radiatively driven with iron opacities playing a central role in the acceleration of the wind \citep{nugis_massloss_2002, grafener_hydrodynamic_2005, vink_metallicity_2005}. Although the luminosities and wind driving mechanism of WR stars are similar to those of OB-type stars, their mass-loss rates, inferred from spectra and atmosphere models, are approximately an order of magnitude larger \citep*{nugis_massloss_2000, puls_mass_2008}. However, the CAK theory of radiative driving by \citet*{castor_radiationdriven_1975}, which successfully explains OB star winds, cannot account for WR winds, primarily owing to its simplifying assumption of single scattering. Indeed, \citet[][hereinafter S20]{sander_nature_2020} showed that WR-type mass-loss begins at the onset of multiple scattering in the wind. Further, proximity to the Eddington limit may be a key cause of WR-type mass loss \citep{grafener_eddington_2011}. It has recently been shown that Eddington-limit-induced mass ejections can reproduce both the Humphreys–Davidson limit and observed WR-star numbers \citep{pauli_drastic_2026}. Sharp changes in mass loss associated with iron ionisation and transitions between wind-driving regimes are also found in near-Eddington WNh stars and OBA hypergiants \citep{lefever_near-eddington_2025, bernini-peron_winds_2026}.

The velocity profile of a radiatively driven wind has often taken the form of a $\beta$-law \citep*{pauldrach_radiationdriven_1986}, $v(r)=v_\infty\left(1- R_*/r \right) ^ \beta$, where $r$ is the distance to the star's centre, $v_\infty$ is the terminal wind velocity,
$R_*$ is the hydrostatic stellar surface radius,
and $\beta$ is a free parameter constrained by spectroscopic observations. Whilst the CAK theory predicts $\beta=0.5$, when accounting for the so-called finite-disk correction\footnote{This is the fact that the radiation pushing the material does not originate in a point source but rather in a finite surface.} $\beta\approx0.8$ fits OB star winds best \citep{Friend1986,Puls1996} and $\beta\gtrsim1$ applies to WR winds \citep{lefever_exploring_2023}. Such a $\beta$-law has been used in WR atmosphere models to estimate mass-loss rates, $\dot{M}$ \citep[e.g.][]{nugis_massloss_2002}. Furthermore, empirical $\dot{M}$ as a function of stellar parameters have been formulated from spectroscopic observations \citep*[e.g.][]{hamann_spectral_1995, nugis_massloss_2000, yoon_better_2017} and these have repeatedly been implemented in WR evolution models \citep[e.g.][]{ georgy_grids_2012, grassitelli_diagnostics_2016, liLi_grids_2023, li_population_2024}. These previous efforts demonstrate a WR radius problem \citep{grassitelli_subsonic_2018}: empirical results for the WR radius $R_*$ seem too high, and so the corresponding temperatures $T_*$ are too low, in comparison to where evolutionary models predict them to be. The photosphere of a star, where its effective temperature $T_\mathrm{eff}$ is defined, is typically where the optical depth $\tau\approx2/3$. However, for the optically thick winds of WR stars, this is satisfied in the supersonically outflowing material. Hence the location of the hydrostatic surface as derived from model atmospheres relies on the treatment of wind structure \citep[e.g.][]{groh_evolution_2014}, for example the implementation of a $\beta$-law or hydrodynamically consistent models.

Recent years have seen progress in the theoretical modelling of the hydrodynamic structure of WR atmospheres (e.g. with the PoWR code; \citealt*{grafener_lineblanketed_2002}; \citealt{hamann_temperature_2003, sander_consistent_2015}). A crucial region in hot WR atmospheres at temperatures $T_\mathrm{e}\approx 200\,\mathrm{kK}$ is the hot iron bump, so called owing to the enhanced bound-bound iron opacities. This opacity bump starts to launch the supersonic wind at deeper layers of the atmosphere \citep{grafener_hydrodynamic_2005, sander_driving_2019, sander_nature_2020, lefever_dynamically_2026}. The surface radii derived from hydrodynamically consistent models usually lie close to these deep wind-launching layers, and are generally significantly lower than those derived empirically. Whilst recent evolutionary studies \citep*[e.g.][]{romagnolo_maximum_2024, gormaz-matamala_evolution_2025} have employed the $\dot{M}(\Gamma_\mathrm{e},Z)$ rates of \citetalias{sander_nature_2020}\footnote{$Z$ is the metallicity. $\Gamma_\mathrm{e}$ is the electron-scattering Eddington parameter defined as the ratio of the acceleration owing to free electron scattering relative to gravitational acceleration. $T_\mathrm{eff,crit}$ is the effective temperature at the launching point of the WR star wind.}, the latest $\dot{M}(\Gamma_\mathrm{e},Z,T_\mathrm{eff,crit})$ updates from \citet[][hereinafter S23]{sander_temperature_2023} have not yet been incorporated in stellar models.

The inclusion of this updated temperature dependency in the mass-loss rate and its effects on WR evolution is the focus of this study. We investigate the evolution of high-mass stars covering a range of stellar masses, metallicities and the latest WR wind mass-loss prescriptions for both single and binary stars. In Section\,\ref{sec:method}, we describe the numerical method and physical assumptions of our stellar evolution models. In Section\,\ref{subsec:Stellar evolution}, we present the key results of effects on stellar evolution and  WR star formation. We discuss the ionising energy output in Section\,\ref{subsec:Ionising radiation} and final stellar masses in Section\,\ref{subsec:Remnants}. We present simulated stellar populations and provide a comparison to observations in Section\,\ref{subsec:Pop synth}. The work is summarised and conclusions are drawn in Section\,\ref{sec:conclusion}.

\section{Numerical method}
\label{sec:method}

We use version 15140 of the open-source 1D stellar evolution code \textit{Modules for Experiments in Stellar Astrophysics} \citep[\mesa;][]{paxton_modules_2011, paxton_modules_2013, paxton_modules_2015, paxton_modules_2018, paxton_modules_2019} to evolve non-rotating single and binary stars with three metallicities, $Z = 0.014$, $Z = 0.0056$ and $Z = 0.00224$. These approximate the metallicity of young massive stars in the Milky Way (MW)\footnote{The choice of $Z = 0.014$ follows the Solar calibration of \citet{Asplund2009}, though recent studies find slightly higher values \citep[e.g.][]{Magg2022,Lodders2025}.}, the Large Magellanic Cloud (LMC) and the Small Magellanic Cloud (SMC), respectively. The chosen metallicities are appropriate for comparisons to observations of WR stars in these galaxies. The initial, zero-age main-sequence (ZAMS) masses of the single stars and the primary stars of binaries in our models are $M_\mathrm{ZAMS}=16$, $20$, $23$, $27$, $31$, $36$, $45$, $52$, $60$, $74$, $99$ and $135\, \mathrm{M}_\odot$. The low end of this mass range is motivated by the initial mass limits for WR star formation (for stars in binaries at $Z=0.014$, the most likely conditions to form a WR star) found by \cite{Shenar2020} and \cite{gilkis_landscape_2025}. These were $18\,\mathrm{M}_\odot$ and $16\,\mathrm{M}_\odot$, respectively. Further, our preliminary results for binary stars at $Z=0.014$ indicate that stars with $M_\mathrm{ZAMS}\lesssim 12\,\mathrm{M}_\odot$ do not give rise to a WR phase during their evolution. Stars more massive than $135\,\mathrm{M}_\odot$ are not modelled because of the high uncertainty in the mass loss and evolution of these very massive stars \citep[VMSs; ][]{grafener_mass_2008,Crowther2010,Bestenlehner2020,sabhahit_massloss_2022}.

For binary systems, whose orbits we assume to be circular \citep*[a reasonable assumption for close binary stars;][]{faccioli_postmainsequence_2008} the mass ratios are $q\equiv M_\mathrm{ZAMS,2}/M_\mathrm{ZAMS,1} = 0.5$ and $0.8$, and the initial orbital periods are $P_\mathrm{i}=5$ and $30\,\mathrm{d}$. While the entire parameter space of possible initial conditions has not been covered, we sample contrasting regimes of close binary interaction, and the grid is not intended to provide a comprehensive exploration of binary parameter space. Four different wind mass-loss prescriptions for the WR phase of stellar evolution are used and assessed. The details of these are given in Section\,\ref{subsec:Wind mass loss}.

The physical assumptions underlying the stellar processes being modelled largely follow \cite{gilkis_landscape_2025}, including a prescription for convective core overshooting following \cite{Jermyn2022}.
Binary interactions are included through Roche lobe overflow (RLOF) mass transfer, focusing on the effect of envelope stripping of the primary which could subsequently become a WR star. We are not concerned with the evolution of the accreting secondary, which we model as a point mass.

The stars are evolved from the
ZAMS
until the end of core carbon burning. This captures almost the entirety of the stellar and WR lifetimes without the complexities and uncertainties of modelling CCSNe and/or NS and BH formation at the end of their lives. Because the evolutionary timescales of nuclear burning become increasingly short towards the end of stellar life, the termination point of the simulations should not have a major effect on modelling observed WR stars or estimating their final masses.

\subsection{Wind mass loss} \label{subsec:Wind mass loss}

\begin{figure*}
    \centering
    \begin{tikzpicture}[node distance=2cm and 2cm, every node/.style={font=\small}]
    
    \tikzstyle{start} = [rectangle, minimum width=3cm, minimum height=1cm, text centered, draw=black, fill=red!25]
    \tikzstyle{process} = [rectangle, minimum width=2.4cm, minimum height=1cm, text centered, draw=black, fill=blue!25]
    \tikzstyle{highlight} = [rectangle, minimum width=2.4cm, minimum height=1cm, text centered, draw=red, dashed, fill=none]
    \tikzstyle{end} = [rectangle, rounded corners, minimum width=2.4cm, minimum height=1cm, text centered, draw=black, fill=green!25]
    \tikzstyle{arrow} = [thick,->,>=stealth]
    \tikzstyle{line} = [thick]
    
    \node (start) at (0,0) [start, text width=2.4cm] {Step 2: \\ Evaluate $\dot{M}$};
    
    \node (Mdot_thin) [start, above left=-0.8cm and 0.2cm of start, text width=2.4cm] {Step 1: \\ Compute $\dot{M}_\mathrm{thin}$};
    
    \node (thinHighH) [process, above left=0.5cm and 0.8cm of Mdot_thin] {$\dot{M}_\text{thin} = \dot{M}_\text{VS21}$};
    \node (thinInterpH) [process, left=2.8cm of Mdot_thin, text width=3.1cm] {Interpolate between $\dot{M}_\text{VS21}$ and $\dot{M}_\text{V17}$};
    \node (thinLowH) [process, below left=0.5cm and 0.8cm of Mdot_thin] {$\dot{M}_\text{thin} = \dot{M}_\text{V17}$};
    
    \node (highCond) [highlight, right=1.6cm of start] {$T_\text{WNL,max} < T_\text{eff}$};
    \node (midhighCond) [highlight, below=0.7cm of start, xshift=1.2cm] {$T_\text{cWR,min} \leq T_\text{eff} \leq T_\text{WNL,max}$};
    \node (midCond) [highlight, below=2.2cm of start, xshift=-1.5cm] {$T_\text{jump} \leq T_\text{eff} < T_\text{cWR,min}$};
    \node (coolCond) [highlight, below left=1.1cm and 0.8cm of start] {$T_\text{eff} < T_\text{jump}$};
    
    \node (highOption1) [end, above=0.6cm of highCond, xshift=-0.1cm, text width=3.5cm] {$\dot{M} = \dot{M}_\text{cWR} \equiv \max( \dot{M}_\text{thin}, \dot{M}_\text{S20/23})$};
    \node (highOption2) [end, below=0.6cm of highCond, xshift=0.4cm] {$\dot{M} = \dot{M}_\text{thin}$};
    
    \node (midhighOption) [end, below=0.3cm of midhighCond, xshift=1.3cm, text width=3.1cm] {Interpolate between $\dot{M}_\text{cWR}$ and $\dot{M}_\text{WNL}$};
    
    \node (midOption) [end, below=0.3cm of midCond, text width=3.1cm] {$\dot{M} = \dot{M}_\text{WNL} \equiv \max( \dot{M}_\text{thin}, \dot{M}_\text{Sab})$};
    
    \node (coolOption) [end, below=0.6cm of coolCond, text width=5.2cm, xshift=-3cm] {$\dot{M} = \dot{M}_\mathrm{cool} \equiv \max(\dot{M}_\text{thin},\dot{M}_\text{dJ88}$)};

    \draw [arrow] (Mdot_thin) |- node[above left] {$X_\text{H,s} \geq 0.7$} (thinHighH);
    \draw [arrow] (Mdot_thin) -- node[above] {$0.4 < X_\text{H,s} < 0.7$} (thinInterpH);
    \draw [arrow] (Mdot_thin) |- node[above left] {$X_\text{H,s} \leq 0.4$} (thinLowH);

    \draw [arrow] (start) -- (highCond);
    \draw [arrow] ([xshift=0.5cm]start.south) -- (midhighCond.north);
    \draw [arrow] ([xshift=-1.2cm]start.south) -- ([xshift=0cm]midCond.north);
    \draw [arrow] (start.south west) -- (coolCond.north);
    
    \draw [arrow] ([xshift=1.1cm]highCond.north) -- node[left] {$\log(L/ \mathrm{L}_\odot)>4.5$} ([xshift=1.2cm]highOption1.south);
    \draw [arrow] ([xshift=1.12cm]highCond.south) -- node[left] {$\log(L/ \mathrm{L}_\odot)\leq4.5$} ([xshift=0.71cm]highOption2.north);

    \draw [arrow] ([xshift=0.7cm]midhighCond.south) -- ([xshift=-0.1cm]midhighOption.north);
    
    \draw [arrow] (midCond.south) -- (midOption.north);
    
    \draw [arrow] (coolCond.south west) -- ([xshift=0.6cm]coolOption.north);
    
    \end{tikzpicture}
    \caption{Wind mass-loss prescription used for massive stars at $Z=0.014$, $0.0056$ and $0.00224$. Details of the mass-loss rates $\dot{M}$ and temperatures $T$ are given in the text.}
    \label{fig:Wind flowchart}
\end{figure*}
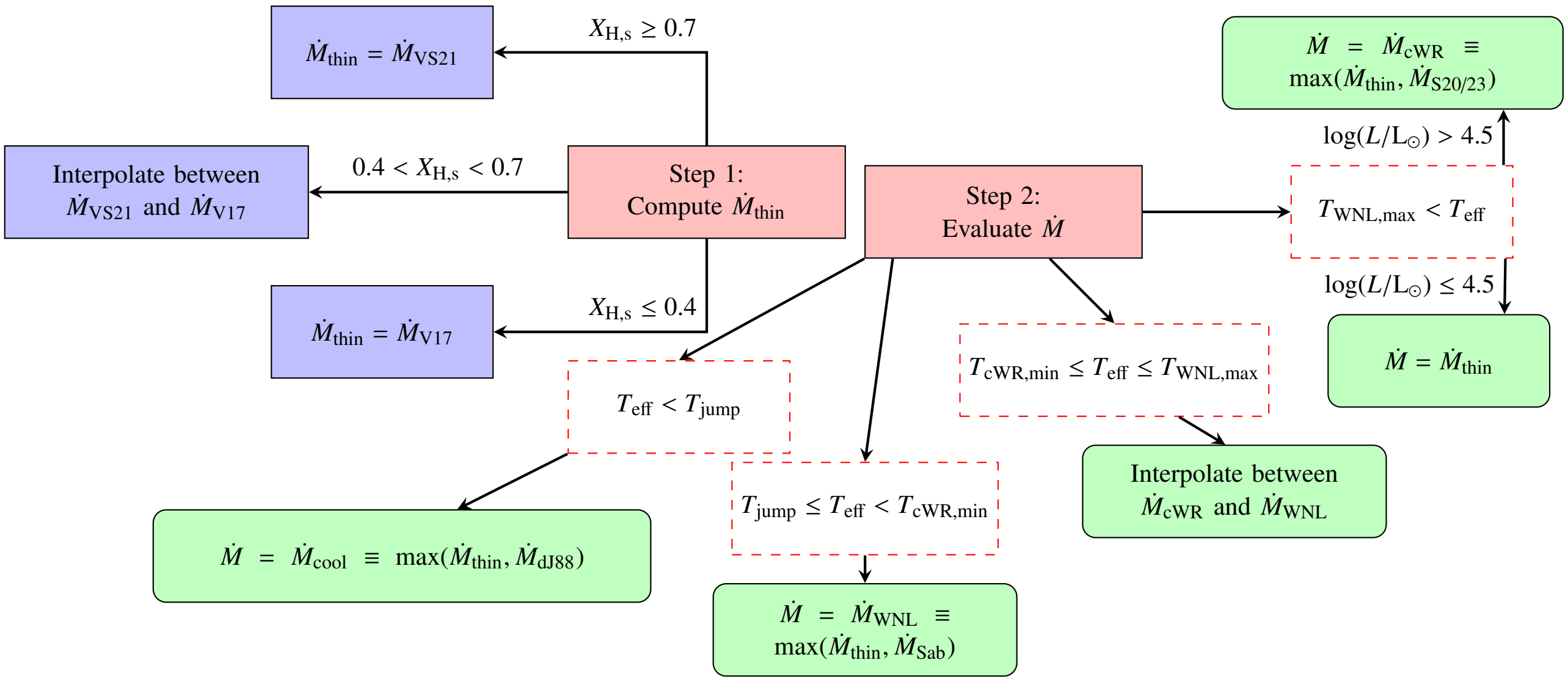
To account for the wide range of luminosities and temperatures encountered during stellar evolution, we apply several mass-loss recipes with empirical rates supported by observations and theoretical rates in the regimes for which they were developed. The approach is broadly similar to that of \cite{gilkis_landscape_2025} and depicted schematically in Fig.\,\ref{fig:Wind flowchart}. An optically thin wind mass-loss rate $\dot{M}_\mathrm{thin}$ is first calculated as $\dot{M}_\mathrm{VS21}$ (following \citealt*{vink_massloss_2001} with the metallicity dependence of \citealt{vink_metallicitydependent_2021}) for hydrogen-rich stars\footnote{Alternative theoretical prescriptions for optically thin winds have been developed by \citet{Bjorklund2023} and \citet*{Krticka2021,Krticka2024,Krticka2025}, while \citet{Pauli2025} recently derived an empirical mass-loss prescription based on observations across Galactic, LMC and SMC metallicities. These prescriptions primarily concern O- and B-type stars, whereas we focus on the subsequent evolution in the optically thick WR-wind regime.} or $\dot{M}_\mathrm{V17}$ (following \citealt{vink_winds_2017}) for helium-rich stars, and interpolating in between.
Four temperature regimes are then considered, as follows.

The hottest regime, $T_\mathrm{eff} > T_\mathrm{WNL,max}$, is for the classical WR stars which have a temperature greater than the maximum expected temperature for the cooler WNL stars, $T_\mathrm{WNL,max}$. Here, $\dot{M}_\mathrm{cWR}$ follows the theoretical recipes of \citetalias{sander_nature_2020} or \citetalias{sander_temperature_2023}. Both have been employed here and their results compared. We note that if the star is not luminous enough, or if the predicted $\dot{M}_\mathrm{S20/23}$ is not high enough, $\dot{M}_\mathrm{thin}$ is used. \citetalias{sander_nature_2020} provide $\dot{M}_\mathrm{S20}(\Gamma_\mathrm{e},Z)$ where, for a star with luminosity $L$ and mass $M$,
\begin{equation} \Gamma_\mathrm{e} = \frac{\sigma_\mathrm{T}}{4\pi c m_\mathrm{H} G} q_\mathrm{ion} \frac{L}{M} = 10^{-4.51}q_\mathrm{ion} \frac{L/ \mathrm{L}_\odot}{M/ \mathrm{M}_\odot} \, \end{equation}
is the Eddington factor. The approximation for the number of free electrons per atomic mass unit, $q_\mathrm{ion}$, from \cite{gilkis_landscape_2025} is used. This allows for surface composition changes and yields $q_\mathrm{ion} \approx 0.5$ for the mostly ionised, H-free atmospheric compositions. The \citetalias{sander_nature_2020} recipe is
\begin{equation} \log_{10} \left( \frac{\dot{M}_\mathrm{S20}}{\mathrm{M}_\odot \, \mathrm{yr}^{-1}} \right) = a \cdot \log_{10} \big[-\log_{10}(1-\Gamma_\mathrm{e})] - \log_{10}(2) \cdot \left(\frac{\Gamma_\mathrm{e,b}}{\Gamma_\mathrm{e}}\right)^c + d , \end{equation}
with the $Z$-dependent coefficients $a, \Gamma_\mathrm{e,b}, c$ and $d$ given by them. \citetalias{sander_temperature_2023} provide $\dot{M}_\mathrm{S23} (\Gamma_\mathrm{e},Z, T_\mathrm{eff,crit})$ where $T_\mathrm{eff,crit}$ is the effective temperature at the launching point of the wind in their WR atmosphere models. This is identified with the effective temperature of the hydrostatic surface $T_\mathrm{eff}$ in our \mesa models.
Although the \textsc{mesa} outer boundary is formally defined at an optical depth of $\tau = 2/3$, this should not be confused with the $\tau = 2/3$ photosphere of an optically thick WR wind, which can lie far above the hydrostatic stellar surface. Rather, the \textsc{mesa} radius approximately corresponds to the hydrostatic boundary underlying the PoWR atmosphere, making $T_\mathrm{eff} \simeq T_\mathrm{eff,crit}$ a reasonable approximation for implementing the \citetalias{sander_temperature_2023} prescription. 

The \citetalias{sander_temperature_2023} recipe is
\begin{equation} \log_{10} \left( \frac{\dot{M}_\mathrm{S23}}{\mathrm{M}_\odot \, \mathrm{yr}^{-1}} \right) = \log_{10} \left( \frac{\dot{M}_\mathrm{S20}}{\mathrm{M}_\odot \, \mathrm{yr}^{-1}} \right) - 6 \log_{10} \left( \frac{T_{\mathrm{eff,crit}}}{141 \, \mathrm{kK}} \right). \end{equation}
This physically-motivated temperature dependency demonstrates that for stars of equal mass, luminosity and composition (and hence $\Gamma_\mathrm{e}$ and $Z$), the cooler star, which is larger and has a lower surface gravity, loses its outer layers more easily and thus has a greater wind mass-loss rate.

The next temperature range covers a transition from the cooler WNL stars (which often have significant amounts of surface H) to the hotter cWR stars, which have a minimum classical WR temperature $T_\mathrm{cWR,min}$. Here, an interpolation between $\dot{M}_\mathrm{cWR}$ and $\dot{M}_\mathrm{WNL}$ ensures a smooth transition in the mass-loss rate. \cite{gilkis_landscape_2025} used $T_\mathrm{cWR,min}=70\,\mathrm{kK}$ for the \citetalias{sander_nature_2020} wind but the models of \citetalias{sander_temperature_2023} either only explored temperatures as low as $T_\mathrm{eff,crit}\approx90\,\mathrm{kK}$ or the winds could not be driven by the hot iron bump at any lower temperatures. Hence $T_\mathrm{cWR,min}=90\,\mathrm{kK}$ is more appropriate for the \citetalias{sander_temperature_2023} wind. Four different WR wind schemes with varying temperature limits are used and assessed. A summary of the parameters is given in Table\,\ref{table:wind prescriptions}. Schemes 1 and 4 are the most suitable implementations of the \citetalias{sander_nature_2020} and \citetalias{sander_temperature_2023} winds, respectively, whilst schemes 2 and 3 act as reference cases for comparison. We do not apply scheme 3 to the binary star models and so it is not included in the simulated stellar populations in Section\,\ref{subsec:Pop synth}.

\begin{table}
    \caption{Parameters of the different WR wind schemes examined in this work. Definitions of the wind types and temperatures are given in the text.}
    \centering
    \begin{tabular}{cccc}
    \hline \hline
    WR wind scheme \# & cWR wind & $T_\mathrm{WNL,max}/\mathrm{kK}$  & $T_\mathrm{cWR,min}/\mathrm{kK}$ \\ \hline
    1           & S20      & 80  & 70 \\
    2           & S23      & 80  & 70 \\ 
    3           & S23      & 90  & 90 \\ 
    4           & S23      & 100 & 90 \\ \hline \hline
    \end{tabular}
    \label{table:wind prescriptions}
\end{table}
The next temperature regime is for the WNL stars, with temperatures between $T_\mathrm{jump}$ and $T_\mathrm{cWR,min}$. The so-called bi-stability jump refers to a change in the wind properties around $21 \lesssim T_\mathrm{jump} / \mathrm{kK} \lesssim 25$ owing to a shift in the ionisation of iron. It is calculated according to \cite{vink_massloss_2001}. In this temperature range, the WNL mass-loss rate $\dot{M}_\mathrm{WNL}$ is found as the maximum of $\dot{M}_\mathrm{thin}$ and $\dot{M}_\mathrm{Sab}$, computed from \cite{sabhahit_massloss_2022} for $Z=0.014$ and $Z=0.0056$, and from \cite{sabhahit_very_2023} for $Z=0.00224$.
The mass-loss properties of the transition from optically thin O-star winds to optically thick WNh winds remain an area of active study. Recent hydrodynamically consistent PoWR calculations have further investigated this transition and provided updated mass-loss predictions as functions of the stellar parameters \citep*{Sabhahit2026a,Sabhahit2026b}.
The precise temperature range over which the WNL and cWR prescriptions should be connected is uncertain, since neither prescription is designed to describe the full transition between these wind regimes. We therefore regard the transition temperatures in Table\,\ref{table:wind prescriptions} as part of the wind implementation rather than fundamental physical boundaries. Schemes 2--4 allow us to assess the sensitivity of the evolution to this choice, with scheme 4 providing a smooth transition while restricting direct application of the \citetalias{sander_temperature_2023} prescription to the temperature range in which hot iron bump is expected to drive winds.

The last temperature range, $T_\mathrm{eff} < T_\mathrm{jump}$, applies to cool stars over a range of evolutionary phases.
The mass-loss rate is taken as the maximum of a cool star rate $\dot{M}_\mathrm{dJ88}$ \citep*[following ][]{dejager_mass_1988} and $\dot{M}_\mathrm{thin}$.

\subsection{WR star properties} \label{subsec:WR properties}

A star is classified as a WR star when its spectrum displays strong emission lines. This occurs when the mass-loss rate is large relative to the size of the star. The strength of the emission features can be characterised by the transformed radius \citep*{schmutz_spectral_1989},
\begin{equation} \label{eq:R_t} R_\mathrm{t}=R_* \left( \frac{v_\infty / 2500\,\mathrm{km}\,\mathrm{s}^{-1}}{\sqrt{D} \dot{M} / 10^{-4}\,\mathrm{M}_\odot\,\mathrm{yr}^{-1}} \right) ^{2/3} \, , \end{equation}
where $D$ is a clumping factor which describes an inhomogeneous wind density \citep{hamann_spectrum_1998}. Physically, the higher density in clumps produces stronger emission than a smooth wind would at the same total mass-loss rate. This term was introduced in spectroscopic analyses to lower mass-loss rates by a factor of $\sqrt{D}$ between $2$ and $3$, and thus aligns models more closely with observations \citep[][]{nugis_massloss_2000}. Modern WR atmosphere codes such as PoWR use a depth-dependent clumping $D(r)$, where it rises from $D(R_*)=1$ to some $D(r\rightarrow\infty)=D_\infty$ (e.g. $D_\infty=50$ by \citetalias{sander_nature_2020}, and $D_\infty=50$, $10$ and $4$ by \citetalias{sander_temperature_2023}).

In our models, WR stars are classified based on the scheme used by \cite{gilkis_landscape_2025}. The main points are summarised here. A star is provisionally classified and assigned $v_\infty$ and $D$ based on its surface mass fractions $X_{i,\mathrm{s}}$ of H, He, N, C and O:
\begin{itemize}
    \item WNh if $X_\mathrm{H,s}\geq0.05$, then $v_\infty = 1000\,\mathrm{km}\,\mathrm{s}^{-1}$, $D=4$;
    \item WC if $X_\mathrm{H,s}<0.05, \quad X_\mathrm{N,s}<X_\mathrm{C,s}, \quad X_\mathrm{C,s}/12 + X_\mathrm{O,s}/16 < X_\mathrm{He,s}/4$, then $v_\infty = 2000\,\mathrm{km}\,\mathrm{s}^{-1}$, $D=10$;
    \item WO if $X_\mathrm{H,s}<0.05, \quad X_\mathrm{N,s}<X_\mathrm{C,s}, \quad X_\mathrm{He,s}/4\leq X_\mathrm{C,s}/12 + X_\mathrm{O,s}/16$, then $v_\infty = 2000\,\mathrm{km}\,\mathrm{s}^{-1}$, $D=10$;
    \item WN otherwise, then $v_\infty = 1600\,\mathrm{km}\,\mathrm{s}^{-1}$, $D=4$.
\end{itemize}
The star is then actually designated as a WR and given its classification if $\log_{10} (R_\mathrm{t}/R_\odot) <1.5$ \citep{Shenar2020}. From its definition in Eq.\,\ref{eq:R_t}, note that stronger winds correspond to higher $\dot{M}$ and thus lower $R_\mathrm{t}$.

\section{Stellar evolution and WR formation}
\label{subsec:Stellar evolution}

\begin{figure*}
    \centering
    \includegraphics[width=\textwidth, trim={0 0 0 24}, clip]{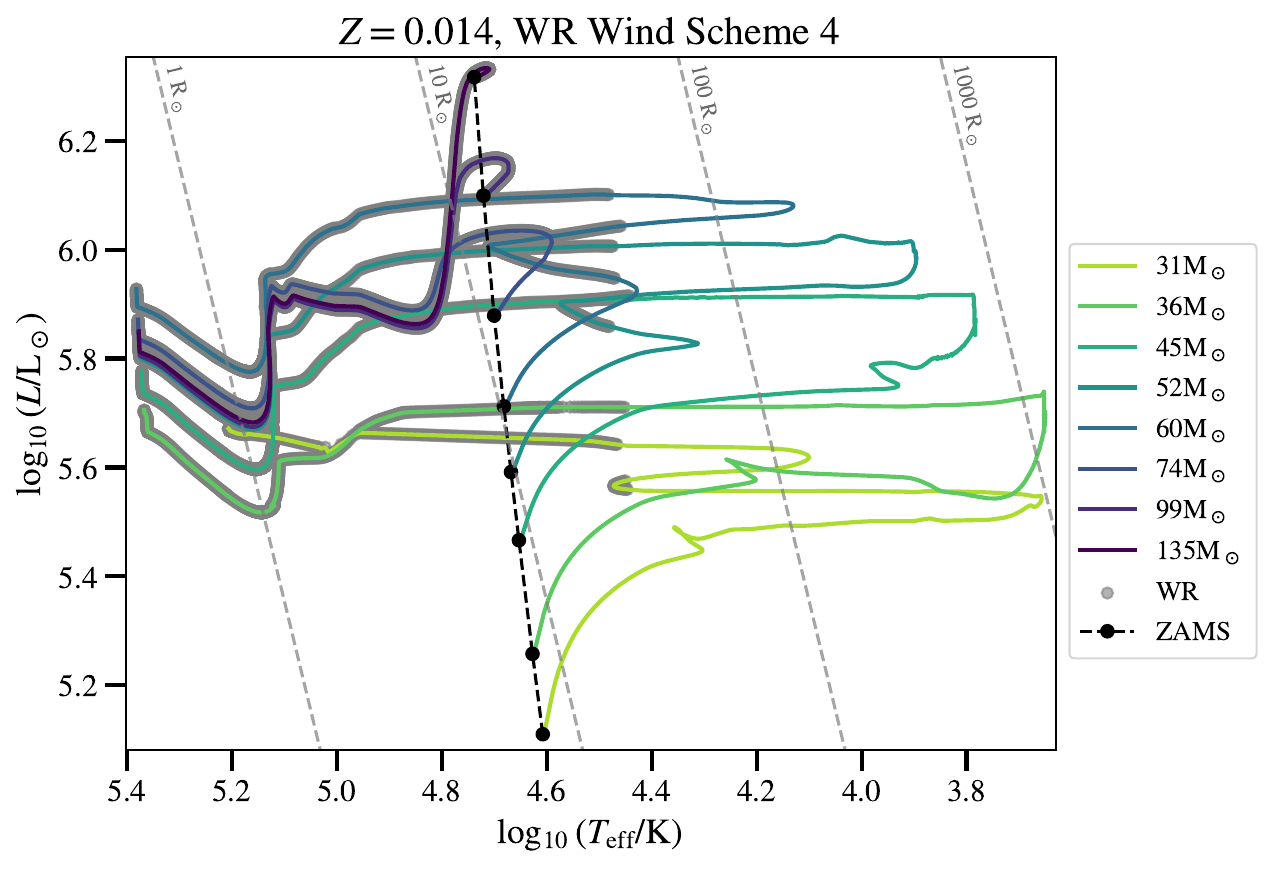}
    \caption{Stellar evolution tracks for single stars with WR wind scheme 4, $Z=0.014$ and ZAMS masses given in the legend. Only stars with an initial mass high enough for the WR phase to be reached are plotted. When a star is predicted to have a WR spectral appearance, it is emphasised with thick grey shading. Each track starts on the ZAMS, denoted by the black dashed line, and lines of constant stellar radius are denoted by the labelled grey dashed lines.}
    \label{fig:HR overview}
\end{figure*}
A sample set of stellar evolution tracks is presented on the Hertzsprung--Russell diagram (HRD) in Fig.\,\ref{fig:HR overview}; these are for $Z=0.014$ and use WR wind scheme 4. At $Z=0.014$, the lower mass stars of this study initially become more luminous and expand as they evolve away from the ZAMS, reaching the cool-supergiant region before subsequently evolving back towards higher effective temperatures. With increasing initial mass, the redward excursion becomes progressively less pronounced: the intermediate-mass models turn back towards the blue at increasingly high effective temperatures, while the most massive stars remain comparatively hot and compact throughout much of their evolution. In particular, the $99$ and $135\,\mathrm{M}_\odot$ stars undergo only modest redward evolution before evolving to higher effective temperatures. This behaviour reflects the increasing importance of stellar winds at high luminosities. These remove the outer hydrogen-rich layers before the stars can expand to the cool-supergiant regime.

The evolutionary track of the $31\,\mathrm{M}$ star in Fig.\,\ref{fig:HR overview} differs from those of more massive stars because it retains some surface hydrogen until the end of their evolution, whereas the higher-mass models become completely hydrogen-free. The sharp increase in luminosity seen towards the end of the higher-mass tracks occurs during core carbon burning and may be associated with structural readjustment during the advanced burning stages, when neutrino losses become increasingly important.

\begin{figure}
    \centering
    \includegraphics[width=0.48\textwidth, trim={0 0 0 24}, clip]{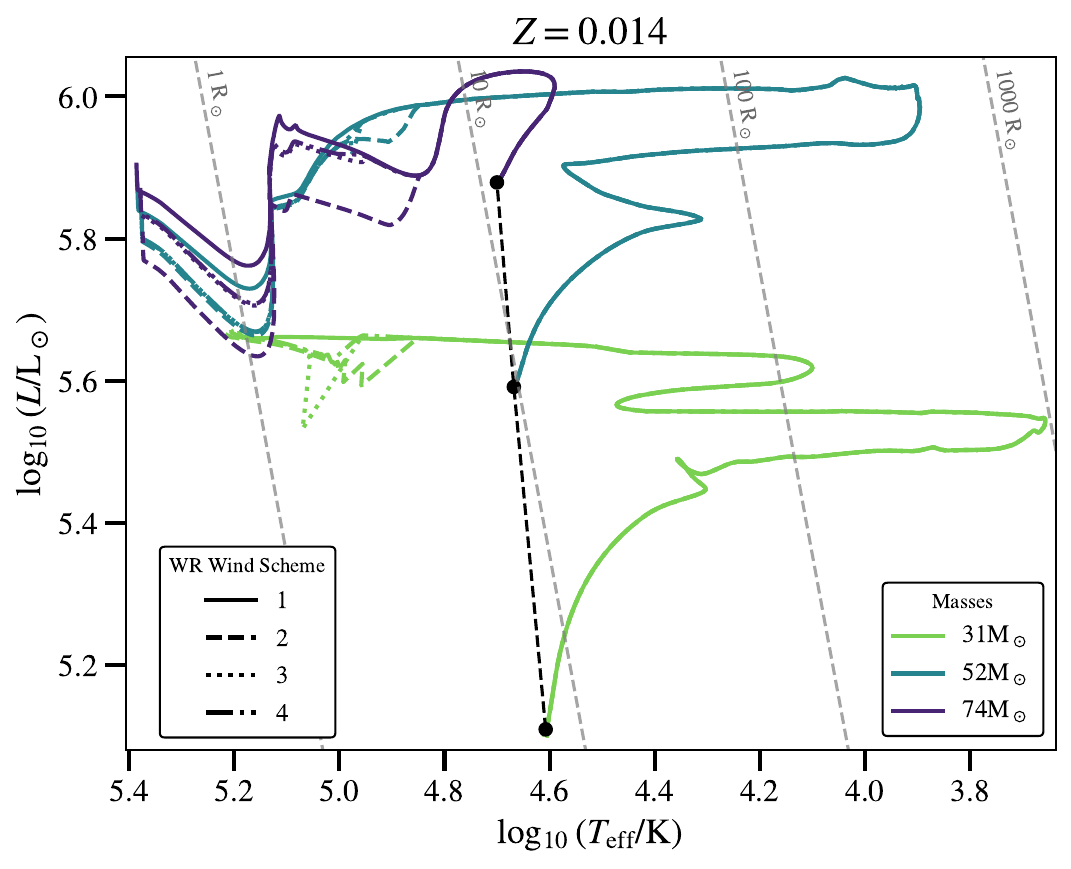}
    \caption{Stellar evolution tracks for single stars at $Z=0.014$, comparing the effect of the different WR wind schemes. The black dashed line is the ZAMS.}
    \label{fig:HR_winddep}
\end{figure}
The effect of the different WR wind schemes on stellar evolution is presented in Fig.\,\ref{fig:HR_winddep}. Single star tracks at $Z=0.014$ are shown. Generally, scheme 2 leads to the greatest mass loss in the early WR phase, leading to lower luminosities later on. Because the \citetalias{sander_temperature_2023} recipe is applied outside of its suggested temperature range for this scheme, these mass-loss rates may be overestimated and can be seen to influence the evolution substantially. We note that the tracks with scheme 3 resemble those of scheme 2 and provide a jagged evolution across the HRD. Therefore, we conclude that scheme 4, with its interpolation allowing for a smooth transition of wind mass-loss rates, is the most appropriate implementation of the \citetalias{sander_temperature_2023} wind. For the highest mass stars, the general evolutionary trend is not sensitive to the choice of wind prescription, though the luminosities at which the stars spend their WR phase vary by up to $25\%$. The intermediate mass stars in this study are least affected by the WR wind scheme choice. They show little variation between the tracks.

\begin{figure}
    \centering
    \includegraphics[width=0.48\textwidth, trim={0 0 0 24}, clip]{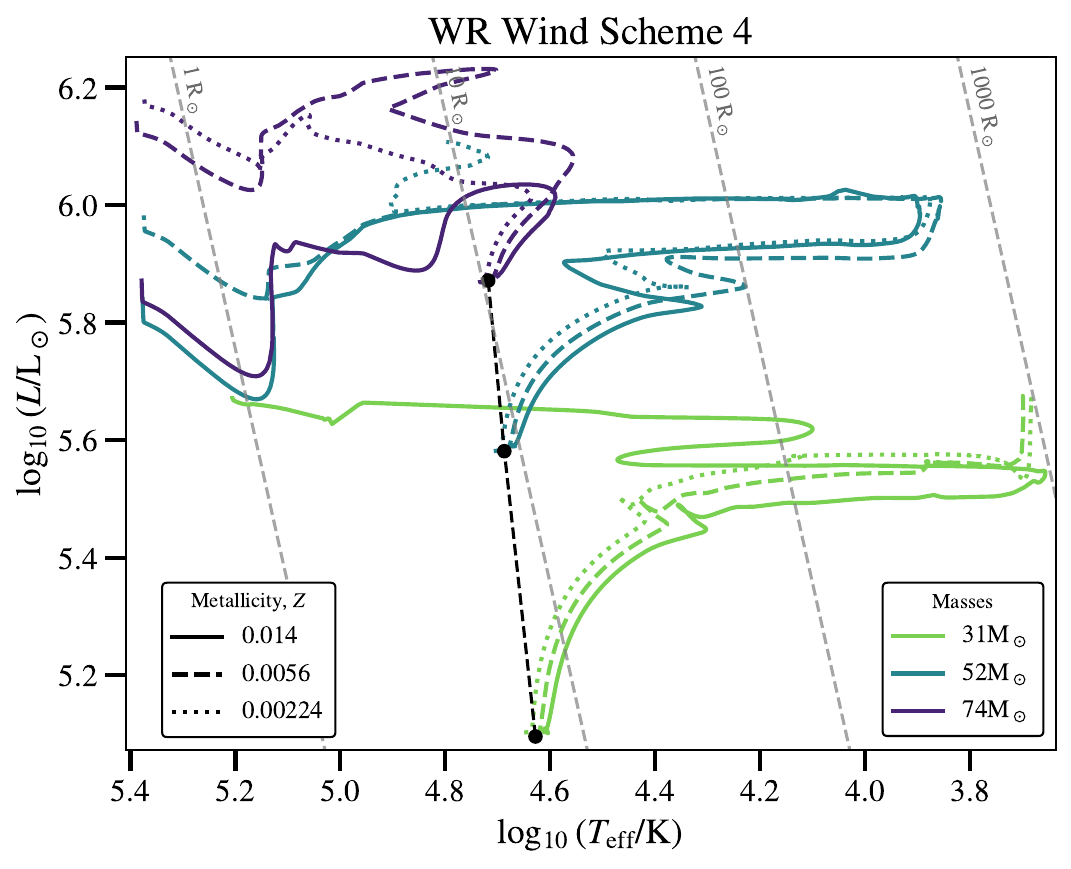}
    \caption{Stellar evolution tracks for single stars with WR wind scheme 4, comparing the effect of the different metallicities. The black dashed line is the $Z=0.0056$ ZAMS. 
    }
    \label{fig:HR_Zdep}
\end{figure}
The effect of metallicity on stellar evolution is presented in Fig.\,\ref{fig:HR_Zdep}. Single star tracks with WR wind scheme 4 are shown. As expected, stars with higher $Z$ have greater mass-loss rates, leading to a more rapid reduction in stellar radius and luminosity. This effect is more pronounced for the higher mass stars because their winds are predominantly radiatively driven and so exhibit a stronger $Z$-dependence. Examination of the $M_\mathrm{ZAMS}=31\,\mathrm{M}_\odot$ tracks gives a hint as to how the metallicity affects which stars  evolve to show WR-like spectral features: below a certain mass threshold at each metallicity, WR stars are not predicted. These thresholds are presented in Table\,\ref{table:M_min for WR}.


Fig.\,\ref{fig:HR_Zdep} also displays an interesting feature for the intermediate-mass stars at $Z=0.00224$ with WR wind scheme 4. Rather than evolving rapidly to the highest temperatures, these stars spend an extended period with $T_\mathrm{eff}$ between $70$ and $90\,\mathrm{kK}$. This behaviour occurs near the transition between the WNL and cWR wind regimes, where the adopted mass-loss prescriptions have qualitatively different dependencies on the stellar parameters. In the \citetalias{sander_temperature_2023} prescription, the cWR mass-loss rate depends steeply on temperature, approximately as $\dot{M}_\mathrm{S23}\propto T_\mathrm{eff}^{-6}$ at fixed $\Gamma_\mathrm{e}$ and $Z$. So evolution towards lower temperatures within the regime in which \citetalias{sander_temperature_2023} is applied strongly increases the cWR mass-loss rate and promotes further stripping of the envelope. However, below $T_\mathrm{cWR,min}=90\,\mathrm{kK}$, the WNL prescription of \citet{sabhahit_very_2023} has no similarly strong explicit temperature dependence, with the enhanced wind rate instead depending steeply on luminosity and mass. The transition between these two wind regimes can modify the response of the star to further expansion or contraction and slow its evolution in effective temperature. As also found for enhanced WNh winds by \citet{sabhahit_very_2023}, strong mass loss governed primarily by the luminosity-to-mass ratio can lead to extended evolution over a relatively narrow temperature range. Consequently, stars near the WR formation threshold can spend a substantial fraction of their WR lifetime at $T_\mathrm{eff}$ between $70$ and $90\,\mathrm{kK}$, increasing their probability of being observed there. This could help explain the characteristic temperature distribution of WR stars in the SMC (Section\,\ref{subsec:Pop synth}).

\begin{figure}
    \centering
    \includegraphics[width=0.48\textwidth, trim={0 0 0 26}, clip]{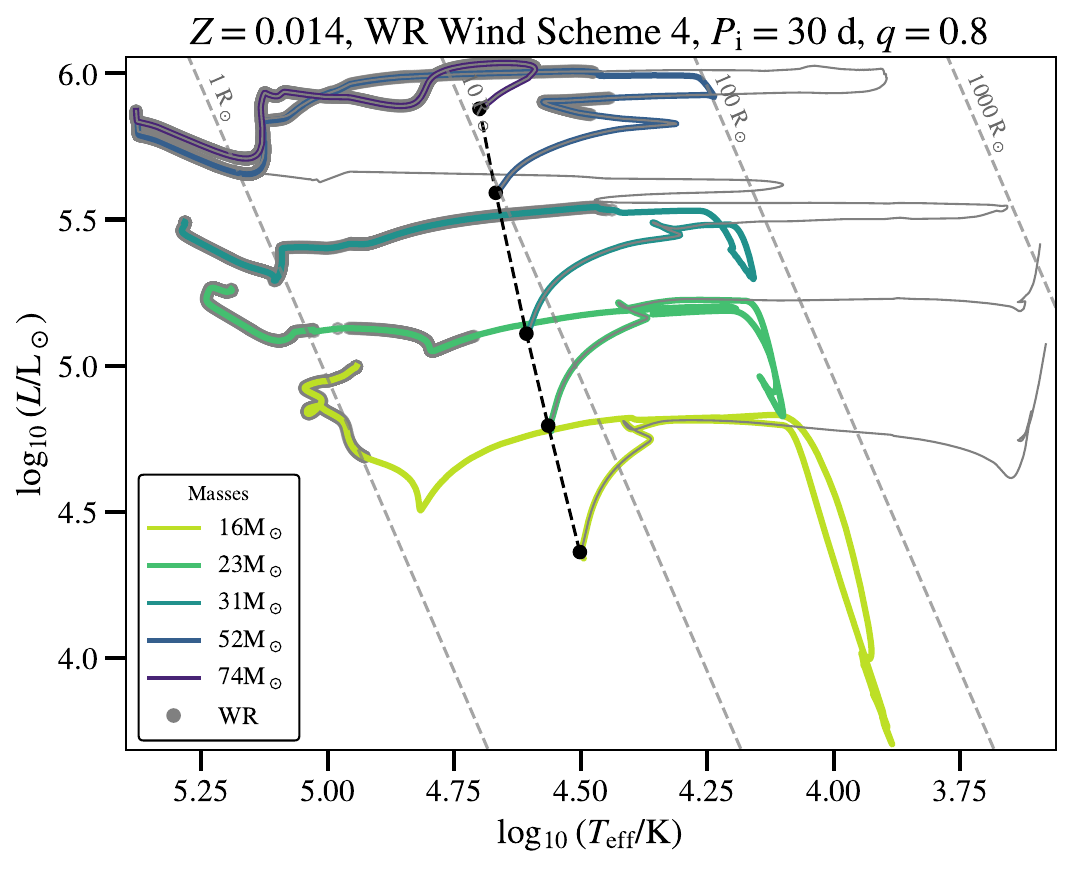}
    \caption{Stellar evolution tracks at $Z=0.014$ for the primary stars in binaries compared to single stars of the same initial mass, denoted by thin grey lines, with initial period $P_\mathrm{i} = 30\,\mathrm{d}$ and mass ratio $q=0.8$.}
    \label{fig:HR_bindep}
\end{figure}
A comparison between single and binary star evolution is presented in Fig.\,\ref{fig:HR_bindep}, with initial conditions similar to those in Fig.\,\ref{fig:HR overview}. For the highest masses, the single-star and binary tracks overlap as the stellar wind is sufficiently strong to strip away the envelope without requiring a companion. At lower masses, the tracks diverge when the stars expand enough to interact. The envelope stripping that ensues for the primary star causes it to lose its outer hydrogen and evolve blueward on the HRD, allowing the WR phase to be reached
for lower masses.

The formation of WR stars is a $Z$-dependent phenomenon largely owing to the weaker stellar winds at lower $Z$. Of interest is the lowest initial mass (denoted by $M_\mathrm{single}^\mathrm{WR}(Z)$ for single stars and $M_\mathrm{binary}^\mathrm{WR}(Z, P_\mathrm{i})$ for the primary stars in binaries) for which the WR phase is predicted to form.
The results for single and binary stars are presented in Table\,\ref{table:M_min for WR}. While the overall metallicity dependence is consistent with previous works \citep{Shenar2020}, we find notable differences in the inferred minimum initial masses. In particular, at $Z=0.00224$ binary interaction can produce a WR star from a primary with an initial mass as low as $23\,M_\odot$ for $P_{\rm i}=5\,\mathrm{d}$. This is more than $10\,M_\odot$ below the corresponding threshold inferred by \citet{Shenar2020}. These objects are relatively low-mass WR stars that emerge only during the late stages of their evolution, which likely contributes to the difference between the inferred thresholds.

\begin{table}
    \caption{Minimum initial masses for which we find a WR phase to exist during the star's evolution as a function of metallicity, the presence of a binary and the binary orbital period.}
    \centering
    \begin{tabular}{cccc}
    \hline \hline
    \multirow{2}{*}{$Z$} & \multirow{2}{*}{$M_\mathrm{single}^\mathrm{WR} / \mathrm{M}_\odot$} & \multicolumn{2}{c}{$M_\mathrm{binary}^\mathrm{WR} / \mathrm{M}_\odot$} \\ \cline{3-4} 
     &  & $P_\mathrm{i}=30 \, \mathrm{d}$ & $P_\mathrm{i}=5 \, \mathrm{d}$ \\ \hline
     0.014   &  31  &  16  &  16  \\
     0.0056  &  36  &  16  &  16  \\
     0.00224 &  45  &  45  &  23  \\ \hline \hline
    \end{tabular}
    \label{table:M_min for WR}
\end{table}
Binary interaction can help form WR stars through envelope stripping of the primary, leading to a lower initial mass threshold for them to form. Whilst it is commonly assumed that binary interaction dominates the formation of WR stars at low $Z$, 
\citet{Shenar2020} have demonstrated that this is not necessarily true because both the emergence of WR features and strong wind mass loss depend on metallicity.
At the two higher metallicities considered, $M_{\rm binary}^{\rm WR}$ is insensitive to the initial orbital period. At the lowest metallicity, however, the shorter-period binary has a substantially lower $M_{\rm binary}^{\rm WR}$. This likely reflects differences in the mass-transfer history, although the relationship between earlier stripping and WR formation is non-trivial owing to its effects on the primary’s luminosity, effective temperature, and subsequent wind mass loss.

\section{Ionising radiation}
\label{subsec:Ionising radiation}

Massive stars emit vast amounts of energy into their surroundings owing to their high luminosities. Their high effective temperatures mean that a significant fraction of their emission is in the UV and can thus contribute to the ionisation of their environment. In particular, the key species to consider are neutral hydrogen atoms (\ion{H}{i}) with ionisation energy $E_\mathrm{ionise}=13.6\,\mathrm{eV}$, neutral helium atoms (\ion{He}{i}) with $E_\mathrm{ionise}=24.6 \,\mathrm{eV}$, and singly ionised helium ions (\ion{He}{ii}) with $E_\mathrm{ionise}=54.4\,\mathrm{eV}$ \citep*{mohr_codata_2016}. This is important for our understanding of early-universe reionisation as well as the formation of ionised bubbles around star-forming regions (\citealt*{bianchi_contribution_2001}; \citealt{krause_feedback_2013}).

\begin{figure}
    \centering
    \includegraphics[width=0.48\textwidth, trim={0 0 0 24}, clip]{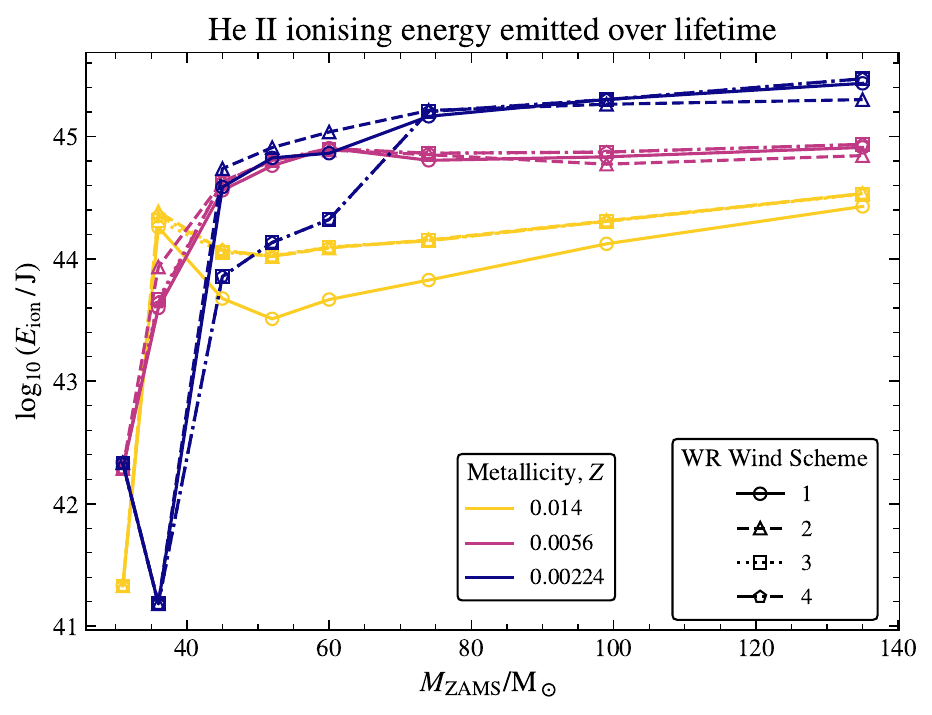}
    \caption{Lifetime-integrated \ion{He}{ii} ionising energy output over stellar lifetimes for single star models, covering the range of metallicities and WR wind schemes. These were approximated with WR atmosphere results from \citealt{sander_temperature_2023}, discussed in detail in the text.}
    \label{fig:Eion_HeII}
\end{figure}
However, despite their high temperatures, WR stars do not necessarily emit a significant number of \ion{He}{ii} ionising photons. The transparency of the wind to photons with energies above $54\, \mathrm{eV}$ depends on the density of the wind and thus the mass-loss rate. In denser winds, \ion{He}{iii} recombines to \ion{He}{ii}, making the atmosphere opaque to these high energy photons (\citealt*{schmutz_theoretical_1992}; \citealt{sander_temperature_2023}). Furthermore, the presence of line blanketing also reduces the emitted ionising fluxes from WR atmospheres \citep{smith_realistic_2002}.

We have used our stellar evolution simulations to provide estimates of the energy outputs of the models. To calculate the total energy emitted over a stellar lifetime, $E_\mathrm{total}$, we simply multiply the stellar luminosity at each time-step with the time-step duration and summed over all time-steps. For the single stars, $E_\mathrm{total}$ is largely independent of $Z$ for $M_\mathrm{ZAMS}<60 \,\mathrm{M}_\odot$, whilst each decrease in metallicity from $Z=0.014$ to $0.0056$ to $0.00224$ yields an increase in $\log_{10} (E_\mathrm{total}/\mathrm{J})$ of about $25\%$ for the higher mass stars (lower $Z$ models generally have higher luminosities). The choice of WR wind scheme shows almost no effect on the total energy output of the stars at all metallicities.

Estimating the ionising energy outputs of the stars requires knowing both their luminosity and how that energy emission is distributed across the electromagnetic spectrum. To this end, the results of \citetalias{sander_temperature_2023} have been applied. For \ion{He}{ii} ionising photons, we use a $Z$-dependent cut-off for their transformed mass-loss rate $\dot{M}_\mathrm{t}$, estimated from Fig.\,17 of \citetalias{sander_temperature_2023} to be 
\begin{equation}
\log_{10} \left(\dot{M}_\mathrm{t}^\mathrm{cut-off} / \left[ \,\mathrm{M}_\odot\,\mathrm{yr}^{-1} \right]\right) = - 4.5 - 0.25 \log_{10}(Z/ \mathrm{Z}_\odot) \, .
\end{equation}
If $\dot{M}_\mathrm{t}$ is larger than this limit, the wind is optically thick. Otherwise, the wind is optically thin. Following \citetalias{sander_temperature_2023}, a temperature $T_{2/3}$ is defined as the effective temperature at the point where an optical depth of $2/3$ is reached. This point typically defines the photosphere and thus dictates the emitted and observed spectra. \citetalias{sander_temperature_2023} provide a fitting formula for $T_{2/3}(\dot{M}_\mathrm{t})$ in their equation\,3. They derived this for their $D_\infty=50$ models so we have shifted $T_{2/3}$ down by $45\%$ to be consistent with the $D_\infty$ assigned to our models (Section\,\ref{subsec:WR properties}). This correction also has the benefit of avoiding any unphysical results when $T_{2/3} > T_\mathrm{eff}$, because we generally expect the effective temperature to decrease outwards.

We define a radiative emission temperature $T_\mathrm{emit}$ for \ion{He}{ii} ionisation. For thin winds $T_\mathrm{emit} = T_\mathrm{eff}$ and for thick winds $T_\mathrm{emit} = T_\mathrm{2/3}$, motivated by considering whether the emergent flux originates from the hydrostatic surface or the outflowing wind. For \ion{H}{i} ionising photons, no such wind density-transparency relationship is observed \citep[Fig.\,21 of][]{sander_nature_2020}, so $T_\mathrm{emit}= T_\mathrm{eff}$ at all times.
A Planck distribution at $T=T_\mathrm{emit}$ is then used to estimate the fraction of energy greater than each ionisation energy limit at each time-step. An estimate of the total ionising energy output over stellar lifetimes, $E_\mathrm{ion}$, is then found by summing over each time-step as above, now weighted by the ionising energy fraction.
We note that the use of a Planck distribution provides only an approximate estimate of the ionising energy output. In particular, departures from local thermodynamic equilibrium (NLTE), line blanketing and absorption within the stellar wind can substantially modify the emergent spectrum, especially at the highest photon energies. Consequently, the quantitative predictions for the ionising energy output should be treated with caution. Our primary aim here is to illustrate how optically thick winds can shift the emergent spectrum towards lower effective temperatures, thereby reducing the ionising output, rather than to provide precise predictions of ionising fluxes. A more quantitative assessment would require detailed atmosphere calculations.

Because $T_\mathrm{emit}=T_\mathrm{eff}$ for \ion{H}{i} ionising photons, the resulting $E_\mathrm{ion,\ion{H}{i}}$ does not differ significantly from $E_\mathrm{total}$, showing the same dependencies on $Z$ and the WR wind scheme used. The increase of $E_\mathrm{ion,\ion{H}{i}}$ with initial stellar mass is more exaggerated than for $E_\mathrm{total}$ because the higher mass stars are on average hotter throughout their lives and so a greater proportion of their emission is above the Lyman limit. Fig.\,\ref{fig:Eion_HeII} presents the total \ion{He}{ii} ionising energy predicted to be emitted by the single stars evolved with the different metallicities and WR wind schemes. A much more complicated picture arises for the less massive stars of this study, for which $E_\mathrm{ion,\ion{He}{ii}}$ is actually smaller at lower $Z$. This effect, noticeable for the single stars with $M_\mathrm{ZAMS}<45\,\mathrm{M}_\odot$ at $Z=0.00224$, arises because these single stars do not reach an evolved WR phase. They die as relatively cool supergiants, having never been hot enough to emit significant amounts of \ion{He}{ii} ionising radiation. In our approximate treatment, $E_\mathrm{ion,\ion{He}{ii}}$ seems to saturate at $M_\mathrm{ZAMS}\gtrsim80\,\mathrm{M}_\odot$, reflecting the suppression of the emergent high-energy radiation by the denser winds of more massive stars. Similar results arise for the binary stars, where $E_\mathrm{ion,\ion{He}{ii}}$ shows a complicated dependency on the various stellar, binary and wind parameters.

\section{Final masses}
\label{subsec:Remnants}

We consider the effects of the WR wind prescription on the final stellar mass.
The final mass of a WR star significantly affects the appearance of a supernova if the star explodes \citep[e.g.][]{GalYam2022}, or the mass of a BH if there is no supernova \citep[e.g.][]{Gilkis2026}.

Having gone through successive stages of nuclear burning and mass loss in winds and binary interaction, our stars finish at the end of core carbon burning with a final stellar mass $M_\mathrm{final}$. The short evolutionary timescales at the end of the evolution mean that the final masses in our models are approximately equal to the pre-core-collapse mass $M_\mathrm{pre-CC}\approx M_\mathrm{final}$. Any further wind mass loss is expected to be negligible. However, eruptions can occur in the remaining months or years before core-collapse \citep{humphreys_luminous_1994, jiang_outbursts_2018, cheng_model_2024}. 

\begin{figure}
    \centering
    \includegraphics[width=0.47\textwidth, trim={0 0 0 24}, clip]{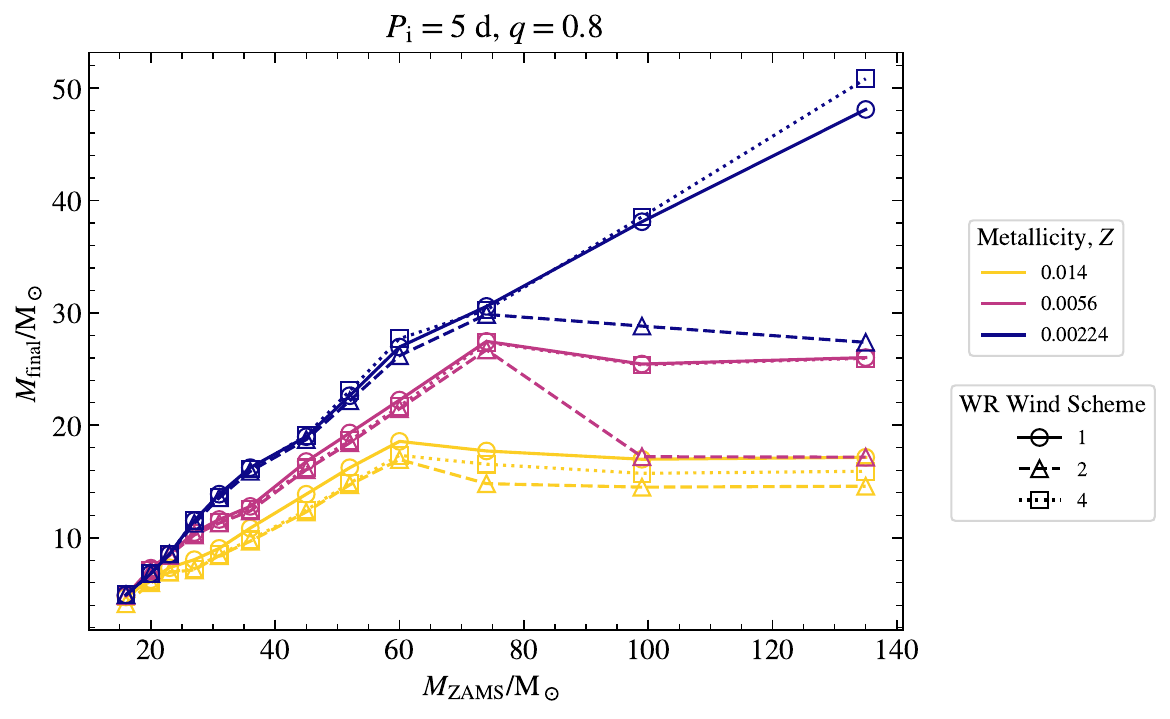}
    \caption{Final mass as a function of initial mass of the stellar evolution models. Shown here are the results for the primary star of binary systems with an initial orbital period of $P_\mathrm{i}=5\,\mathrm{d}$ and a mass ratio of $q=0.8$. These depend on the choice of metallicity (different colours) and WR wind scheme (different markers and line styles).}
    \label{fig:final_zams}
\end{figure}
Fig.\,\ref{fig:final_zams} shows $M_\mathrm{final}$ as a function of $M_\mathrm{ZAMS}$ for the primary star in binaries with $P_\mathrm{i}=5\,\mathrm{d}$ and $q=0.8$. As discussed, the stronger winds at high $Z$ cause much greater mass loss, leading to significantly smaller final masses. In particular, for $Z=0.014$, the final mass saturates at about $15\,\mathrm{M}_\odot$ for $M_\mathrm{ZAMS}\geq 60\,\mathrm{M}_\odot$. 
This is a result of the post-MS evolution of these high-mass stars being similar to that of their single-star counterparts (Fig.\,\ref{fig:HR overview}), because their strong winds dominate the envelope stripping even in binary systems.
This low $M_\mathrm{final}$ saturation at $Z=0.014$ disagrees qualitatively with previous works (e.g. \citealt*{woosley_birth_2020}; \citealt{winch_black_2025}). However the models of \cite{woosley_birth_2020} start as naked He stars, having skipped the WNL phase entirely, so this discrepancy could result from the high uncertainty in the mass loss of VMSs. While \cite{winch_black_2025} show that rapid rotation can substantially modify $M_\mathrm{final}$ through rotationally enhanced and mechanical mass loss, their models span $Z=0.1$ to $0.001\,Z_\odot$, predominantly below the metallicities considered here, and have initial rotation rates of $\Omega = 0.4$ to $0.8 \Omega_\mathrm{crit}$, whereas our models are non-rotating. Further, $M_\mathrm{final}$ is lowest at all $Z$ for WR wind scheme 2. Again this is likely due to the misapplication of the \citetalias{sander_temperature_2023} recipe for cWR stars. So it is promising to see quantitative agreement between the results of WR wind schemes 1 and 4.

Observed BH masses provide an additional constraint on the final stellar masses predicted by our models. In the Milky Way, the mass of the BH in Cygnus X-1 was estimated to be approximately $21\,\mathrm{M}_\odot$ by \citet{MillerJones2021}, exceeding the saturation mass of approximately $15\,\mathrm{M}_\odot$ found in our solar-metallicity models. However, a subsequent analysis by \citet{Ramachandran2025} revised this estimate to $12$--$18\,\mathrm{M}_\odot$, overlapping the range predicted by our solar-metallicity models. In the LMC, the BH masses inferred for LMC X-1 ($10.91\pm1.4\,\mathrm{M}_\odot$; \citealt{Orosz2009}) and VFTS 243 ($10.1\pm2.0\,\mathrm{M}_\odot$; \citealt{Shenar2022}) are below the final stellar masses predicted for the more massive progenitors in our models. These comparisons are indicative rather than definitive, because the final stellar mass need not equal the BH mass and the observed systems may have followed different evolutionary pathways.

\section{Probability-weighted stellar populations} \label{subsec:Pop synth}

We use our models to construct populations of WR stars at MW, LMC and SMC metallicities. A probability is assigned to each model based on estimated star formation statistics. The approach is similar to that of \cite{gilkis_landscape_2025}. We use an initial mass function (IMF) of $\mathrm{d} N / \mathrm{d} M \propto M^{-2.3}$ suitable for the high-mass stars investigated in this study \citep{Salpeter1955}. Each discrete model mass $M_i$ represents the interval $[M_{i,-} , M_{i,+}]$ bounded by the midpoints between adjacent grid masses and its weight is therefore
\begin{equation} \label{eq:w_M,i} w_{M,i} \propto \int_{M_{i,-}}^{M_{i,+}}M^{-2.3} \mathrm{d} M \, . \end{equation}
The population is restricted to $16\leq M/M_\odot\leq135$, which define the outer boundaries of the lowest- and highest-mass bins, respectively. For binaries in the parameter ranges considered here, we assume a flat distribution in the mass ratio, $\mathrm{d}N/\mathrm{d}q =\,\text{const.}$ \citep{shenar_tarantula_2022}, and an initial-period distribution $\mathrm{d}N/\mathrm{d}(\log P) \propto (\log P)^{-0.55}$ \citep{sana_binary_2012}. We adopt $0.1\leq q\leq1$ and $1\leq P/{\rm d}\leq10^4$ as the limits of these distributions. The sampled values $q=0.5,0.8$ and $P=5,30\,{\rm d}$ divide these ranges at the midpoints between adjacent grid points (in $q$ and $\log P$, respectively), with $w_q$ and $w_P$ obtained by integrating the corresponding distributions over these intervals, just as in Eq.\,\ref{eq:w_M,i}. We assume a binary fraction $f_\mathrm{bin}=0.7$ \citep{sana_binary_2012}. The weight of each track is then
\begin{equation} \label{eq:w_track} w_\mathrm{track} = \begin{cases} 
      (1-f_\mathrm{bin}) w_M, & \text{ single star,} \\
      f_\mathrm{bin} w_M w_q w_P, & \text{ binary star.}
   \end{cases} \end{equation}
Finally, each time-step is assigned a weight $w' = w_\mathrm{track} \Delta t$, after which the weights are normalised over the complete population:
\begin{equation} \label{eq:w_final} w = \frac{w_\mathrm{track}\Delta t}{\Sigma_j w_\mathrm{track,  j}\Delta t_j} \, . \end{equation}
The resulting population thus accounts for both the expected formation probability of each system and the time spent at each stage of its evolution. We have also restricted the weights to only include time-steps when the star is considered to be a WR star as outlined in Section\,\ref{subsec:WR properties}. Separate populations are constructed for each metallicity and WR wind scheme and are examined on the HRD and $\log\dot{M}$--$\log L$ plane during the WR phase.



\begin{figure}
    \centering
    \begin{subfigure}{0.47\textwidth}   
        \includegraphics[width=1\textwidth, trim={0 70 0 22}, clip]{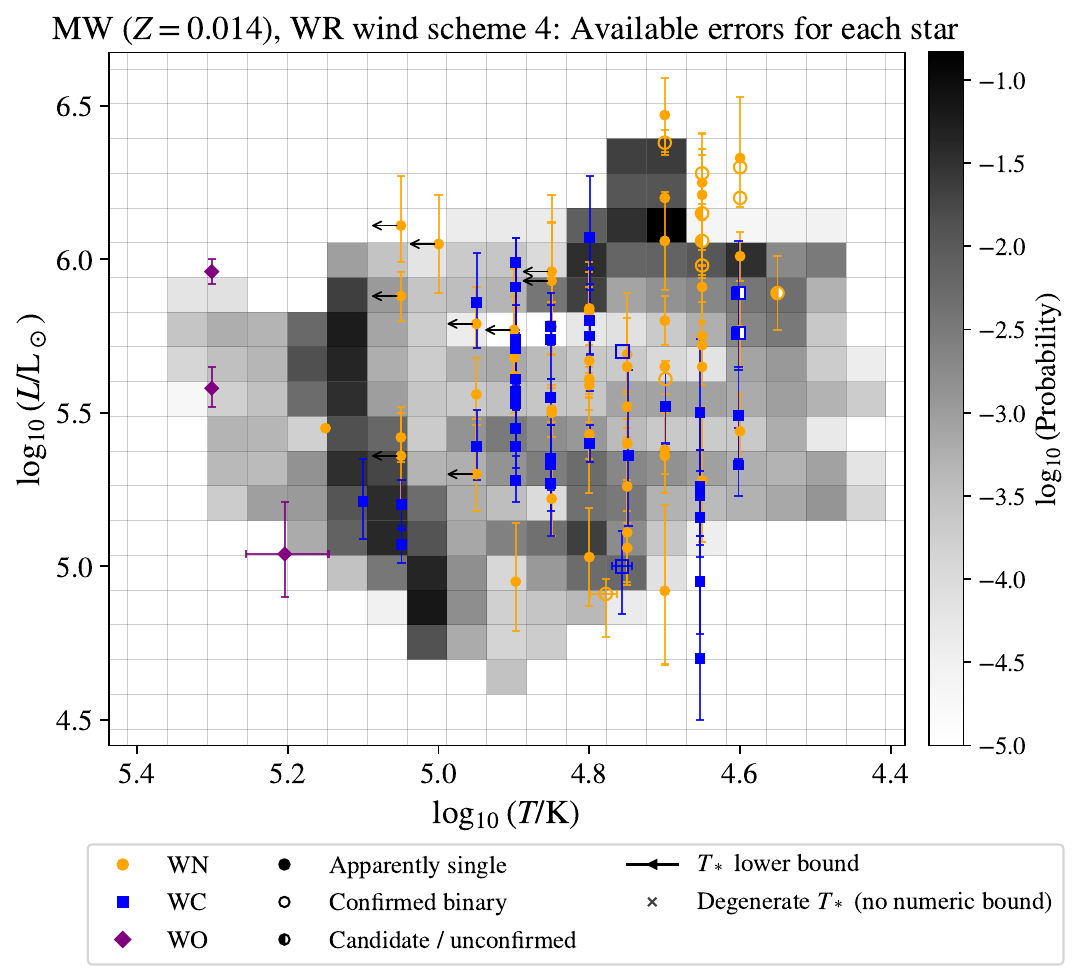}
        \label{fig:popsynth_HR_a}
    \end{subfigure}\\
    \begin{subfigure}{0.47\textwidth}   
        \includegraphics[width=1\textwidth, trim={0 70 0 22}, clip]{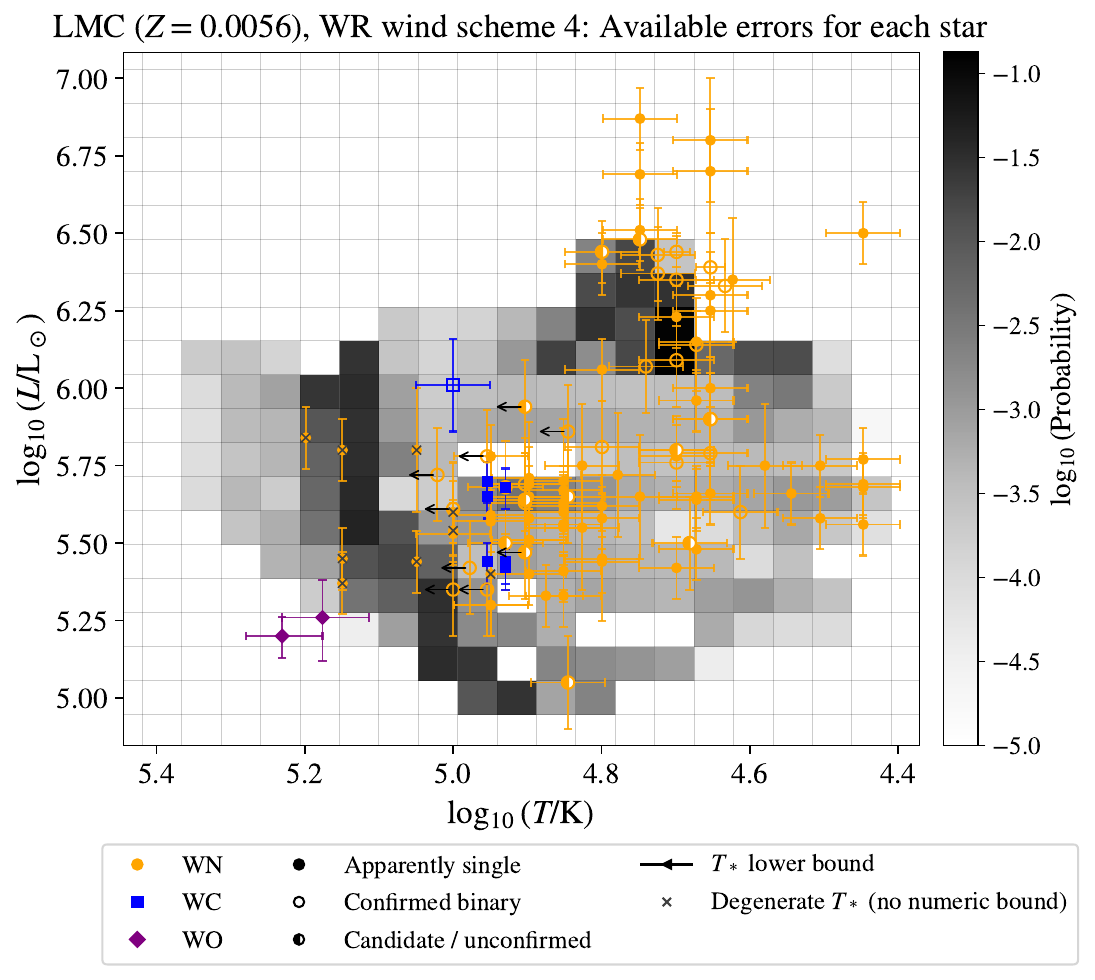}
        \label{fig:popsynth_HR_b}
    \end{subfigure}\\
    \begin{subfigure}{0.47\textwidth}   
        \includegraphics[width=1\textwidth, trim={0 0 0 22}, clip]{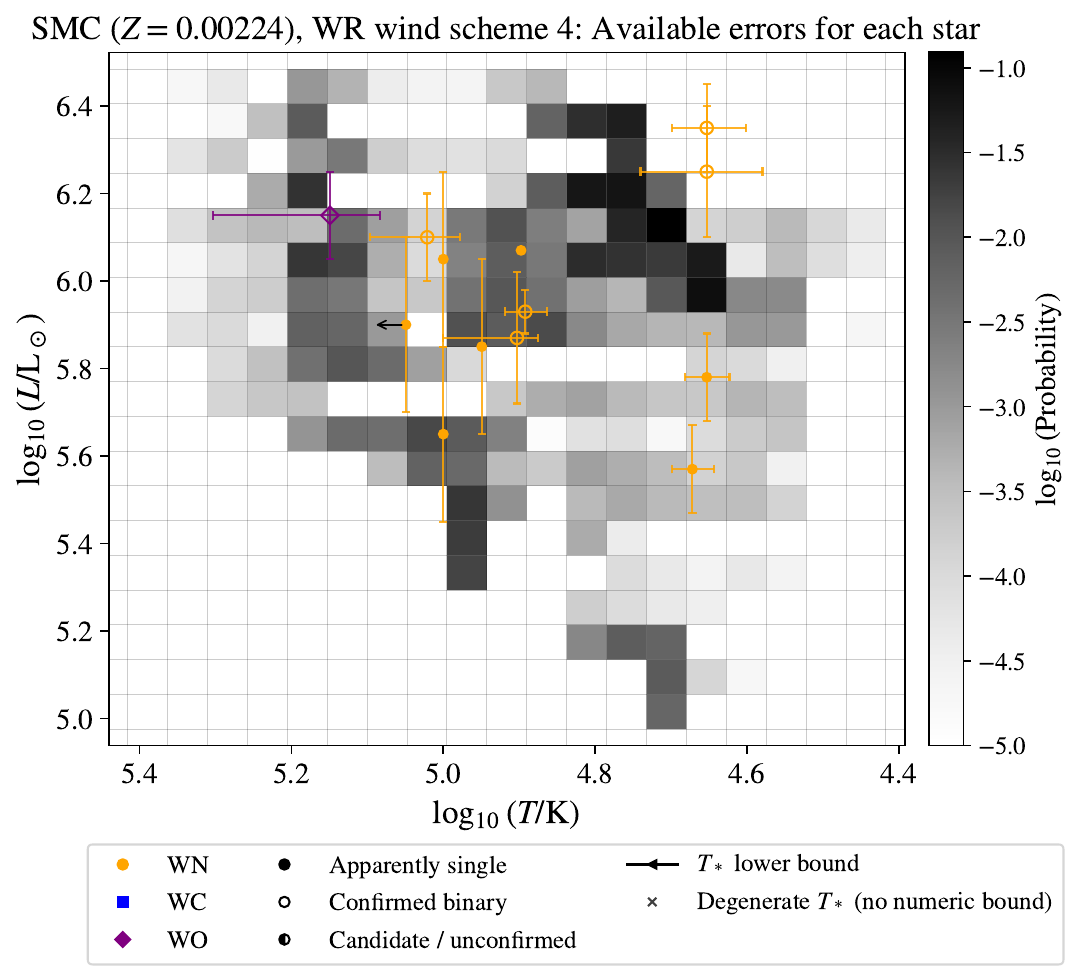}
        \label{fig:popsynth_HR_c}
    \end{subfigure}\\
    \caption{Probability map of our WR stars with WR wind scheme 4, for $Z=0.014$ (top), $Z=0.0056$ (middle) and $Z=0.00224$ (bottom), on the HRD with observed Local-Group WR stars overlaid. The observations are coloured based on their WR classification. Closed symbols denote single WR stars and open symbols denote WR stars in binary systems.
    }
    \label{fig:popsynth_HR_Z0.014_wind4}
\end{figure}
Observations of Galactic single WN stars \citep{hamann_galactic_2019}, Galactic single WC and WO stars \citep{sander_galactic_2019}, a Galactic binary WN5h star \citep{muller-horn_galactic_2026}, LMC WN stars \citep{hainich_wolfrayet_2014}, LMC binary WN stars \citep{shenar_wolf_2019, tehrani_weighing_2019, shenar_wolf-rayet_2020}, SMC single WN stars \citep{hainich_wolfrayet_2015}, SMC binary WN stars \citep{shenar_wolfrayet_2016} and WC and WO stars in all three populations (\citealt{demarco_velorum_2000, dessart_quantitative_2000, crowther_stellar_2002, tramper_massive_2015}; \citealt*{tramper_new_2016}) have been analysed. Updated values for the SMC WN binary star AB 6 and the LMC WN binary star R 144, from \citet{shenar_shortest-period_2018} and \citet{shenar_tarantula_2021}, respectively, have been used. Characterising observed WR properties requires atmospheric and spectral modelling, for which these studies have used PoWR \citep{grafener_lineblanketed_2002, hamann_temperature_2003, sander_consistent_2015} or CMFGEN \citep{hillier_treatment_1998}. These typically employ a $\beta$-law velocity profile for the supersonic part of the wind. Some use a constant clumping factor $D$ and some use a depth-dependent $D(r)$.



Fig.\,\ref{fig:popsynth_HR_Z0.014_wind4} compares the probability-weighted model populations with observed WR stars in the three metallicity environments.
The models reproduce much of the observed WR luminosity range and predict concentrations of WN and WC stars towards the lower end of their expected temperature ranges, although substantial probability remains at higher temperatures than inferred observationally.
The model's probability peak at around $100$ to $140\,\mathrm{kK}$ may partly reflect an overestimate of the effective temperatures, because recent work has shown that detailed atmosphere boundary conditions can substantially reduce WR effective temperatures at solar metallicity without significantly altering their internal structure \citep*{voje_wolf-rayet_2026}. Conversely, observationally inferred WR temperatures may be underestimated, particularly for stars with optically thick winds, owing to uncertainties in the atmospheric and wind structure. Further, the hottest and most evolved WO stars with $T_\mathrm{eff}\gtrsim 140\,\mathrm{kK}$ are both theoretically predicted and observed to be a rare but important class of WR stars. 

The LMC ($Z=0.0056$) population shows some agreement with our simulations, although the most luminous observed WR stars, reaching $\log_{10}(L/\mathrm{L}_\odot)\approx6.8$, are likely core-hydrogen-burning stars with initial masses beyond the $135\,\mathrm{M}_\odot$ upper limit of our models.
At $Z=0.00224$, a slight gap in luminosity in the range
$5.7 \lesssim \log_{10}(L/\mathrm{L}_\odot) \lesssim 5.8$
is predicted by our models. This gap might be a result of a limited number of simulations and larger model grids could provide a more accurate predicted distribution.


Under WR wind scheme 4 at $Z=0.00224$, the intermediate mass WR stars spend extended periods at $70$ to $90\,\mathrm{kK}$, as noted in Section\,\ref{subsec:Stellar evolution}. This is particularly pronounced for stars near the WR star formation threshold, namely those at low $Z$ and at (relatively) low mass in binary systems. Almost all of the observed WR stars in the SMC appear in this temperature range. This has previously been difficult to reproduce theoretically, and our new models with the most up-to-date WR wind prescription alleviate this discrepancy. We note also that the observed temperatures are more reliable for stars in the SMC because their winds are not as optically thick.


\begin{figure}
    \centering
    \includegraphics[width=0.48\textwidth, trim={0 0 0 22}, clip]{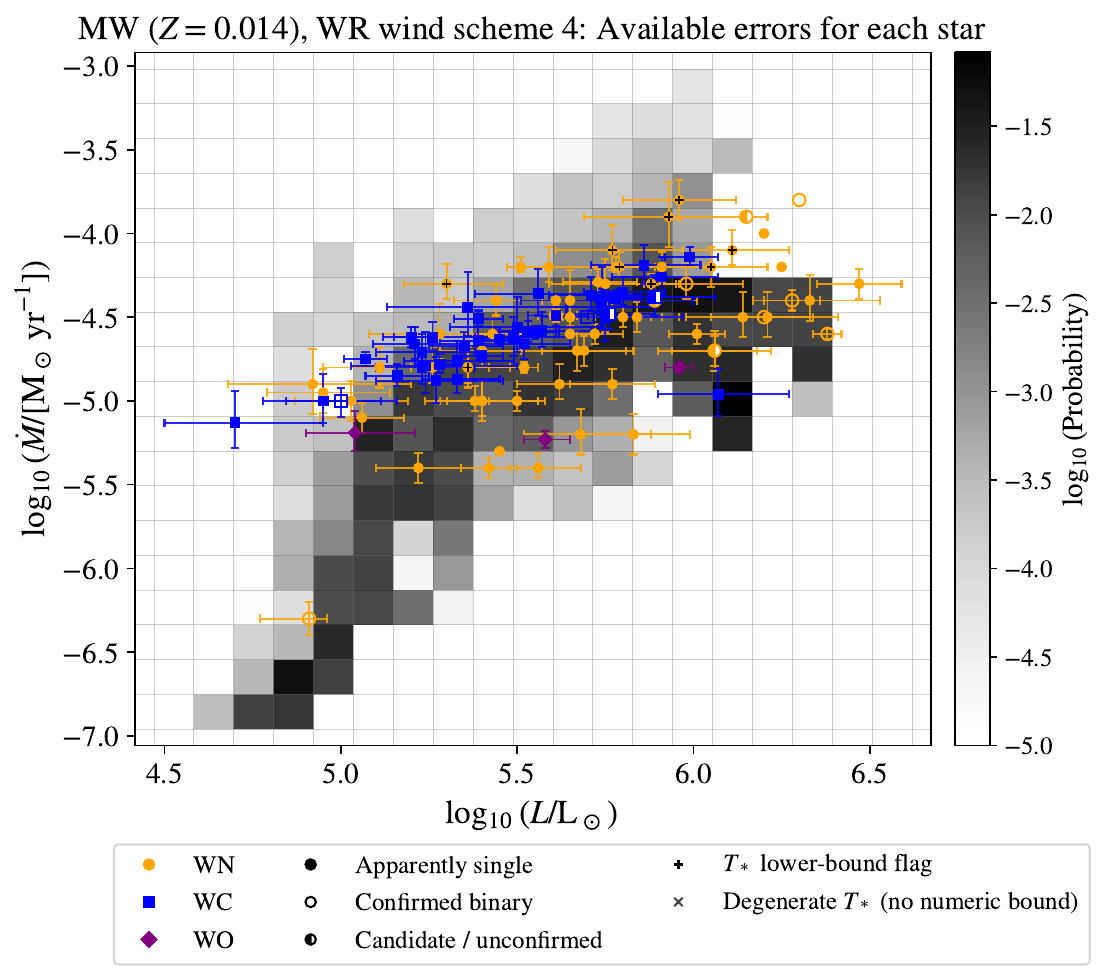}
    \caption{Probability map of simulated WR stars, for $Z=0.014$ and WR wind scheme 4, on the $\log\dot{M}$--$\log L$ plane with observed Galactic WR stars overlaid, similar in construction to Fig.\,\ref{fig:popsynth_HR_Z0.014_wind4}.}
    \label{fig:popsynth_MdotL_wind4}
\end{figure}
Fig.\,\ref{fig:popsynth_MdotL_wind4} is constructed similarly to Fig.\,\ref{fig:popsynth_HR_Z0.014_wind4} but displays the simulated and observed Galactic populations on the $\log\dot{M}$--$\log L$ plane.
While the models span much of the observed range of mass-loss rates and luminosities, they also predict a population extending to lower mass-loss rates than those of most observed WR stars. The recent identification of a low-luminosity WR star by \citet{muller-horn_galactic_2026} provides evidence that such objects exist, although their prevalence remains uncertain. This low-mass-loss population may represent an extension of the observed WR distribution that has largely escaped detection, or it may partly reflect the classification of model stars as WR stars when their winds are insufficiently strong to produce characteristic WR spectra.
\section{Summary and conclusions}
\label{sec:conclusion}

We used the stellar evolution code \mesa to model single and binary stars at Milky Way, LMC and SMC metallicities, incorporating four wind prescriptions, including the temperature-dependent mass-loss recipe recently developed by \cite{sander_temperature_2023}. These models provide an opportunity to link theoretical wind-driving physics to observable WR properties across different environments.



WR formation thresholds depend on metallicity and binary interaction but are largely insensitive to the wind prescription. Binary mass transfer enables WR star formation from lower initial masses than single-star evolution.
Stellar evolution differences emerge near the WR formation mass limit. With our wind scheme 4 (including the \citealt{sander_temperature_2023} temperature dependence) some stars stall between $70$ and $90\,\mathrm{kK}$, matching observed SMC WR temperatures that earlier models struggled to reproduce.
Total and \ion{H}{i}-ionising fluxes show little dependence on wind prescription. \ion{He}{ii}-ionising fluxes saturate at high masses because thick winds absorb high-energy photons.
The population synthesis spans much of the observed WR luminosity and mass-loss-rate ranges, while some discrepancies remain in the effective-temperature distribution.

Incorporating the updated temperature-dependent WR wind \citep{sander_temperature_2023} in stellar evolution models improves agreement with observed WR temperature distributions, especially in low-metallicity environments, while preserving consistency in predicted luminosities, ionising fluxes and mass-loss rates. This represents a step forward in linking theoretical wind physics to WR population properties, though some aspects, such as binary interactions and eruptive mass loss, remain key uncertainties.
Addressing these open issues will be key to fully understanding how WR stars form, evolve and contribute to the feedback and chemical enrichment of galaxies.

\section*{Acknowledgments}

AG acknowledges funding from STFC grant UKRI4122.
TS acknowledges support from the ERC under the European Union’s Horizon 2020 research and innovation program, grant agreement number 101164755 and from the Israel Science Foundation (ISF) under grant number 2434/24.
CAT thanks Churchill College for his fellowship.

\section*{Data Availability Statement}

The data underlying this article will be made available upon publication.

\bibliographystyle{mnras}
\input{main.bbl} 

    \label{fig:popsynth_HR_Z0.00224}

\label{lastpage}

\end{document}